\documentstyle[12pt]{article}
\begin{document}
\title{Random copolymers at a selective interface: many chains with
excluded volume interactions} \author{Gongwen Peng$^{\dag, \S}$,
Jens-Uwe Sommer$^{\dag}$ and Alexander Blumen$^{\dag}$\\ \small \dag)
Theoretische Polymerphysik, Universit\"at Freiburg, Rheinstr.12,
D--79104 Freiburg, Germany\\ \small \S) Department of Physics and Astronomy, Bowling Green State University, Bowling Green, OH 43403} \maketitle
\begin{abstract}
We investigate numerically, using the bond--fluctuation model, the
adsorption of many random AB--copolymers with excluded volume
interactions at the interface between two solvents.  We find two
regimes, controlled by the total number of polymers. In the first
(dilute) regime, the copolymers near the interface extend parallel to
it, while in the second regime they extend perpendicular to it. The
density at the interface and the density in the bulk depend
differently on the total number of copolymers: In the first regime the
density at the interface increases more rapidly then in the bulk,
whereas the opposite is true in the second regime.  \\
\end{abstract}
\vspace{0cm}
\newpage

\noindent
{\bf I. Introduction}
 
Recently copolymers at selective interfaces
have received much attention~[1--6]. Consider as an example a block
copolymer consisting of a hydrophilic and a hydrophobic part and
situated next to an oil--water interface.  The difference in the
monomers' selectivity (which we denote by $kT\chi$) favors the
localization of the copolymer at the interface, with each block in its
favorable solvent. However, for {\it random} copolymers frustrated
situations may arise, since the chain's connectivity forces some
monomers to stay in their unfavorable solvent.  Garel $et$
$al$~\cite{Garel} have studied the localization transition of an ideal
random chain at a selective interface. In previous
publications~\cite{us_1,us_2} we have extended this approach to single
chain under good solvent conditions, and we have shown that for the
adsorption of random copolymers a simple scaling picture works very
well. The scaling picture is based on the fact that the static
properties of a single adsorbed chain consisting of $N$ monomers may
be understood in terms of blobs. A blob is made of $g$ monomers; hence
in it one of the monomer species is on the average in excess by
$g^{1/2}$. The blob stays in its preferred solvent as long as its
total interface selectivity, $\sim g^{1/2}{\chi}k_BT$, counterbalances its
translational free energy, $k_BT$. This leads to $g\chi^2=C$, where
$C$ is model dependent numerical constant. For our lattice simulation model
$C$ turns out to be larger than 10. Usually this constant $C$  (which
does not play any role in a scaling argumentation) is set to
unity. It 
follows that the number of blobs in the chain, $N/g$, equals
$N\chi^2/C$. 
In fact $N\chi^2$ turns out to be the scaling variable of
the problem, as we have confirmed through Monte Carlo
simulations~\cite{us_1,us_2}. 
In Ref.\cite{us_last} we extended our analysis to asymmetric interface potentials, so
that the chain as a whole prefers one solvent side. This problem leads to two new
critical exponents (as predicted from scaling); we succeeded in evaluating
these exponents based on our simulation data~\cite{us_last}.

The present article is devoted to copolymer systems consisting of many
chains, so that the polymer bulk density $\rho_b$ is significant.  As
we proceed to show, due to their interaction with the interface and
depending on $\rho_b$ the copolymers display (at least) two distinct
regimes. At very low $\rho_b$ the interface is only partially covered
with adsorbed blobs; the shape of the adsorbed chains is rather flat,
since their extension is larger parallel to the interface than
perpendicular to it. In fact, the extension of isolated chains
perpendicular to the interface does not depend on $N$, but only on the
interface selectivity $\chi$~\cite{us_1,us_2}. As $\rho_b$ increases,
the adsorbed, flat chains influence each other through excluded volume
interactions. The chains at the interface form thus a two-dimensional
semi-dilute solution. Increasing $\rho_b$ further leads to an
interface completely saturated with blobs; thus the interface density
$\rho_s$ depends in a complex manner on $\rho_b$.  If the adsorption
is strong enough, i.e. if the free energy of adsorption per chain is
much larger than $kT$, the surface can become saturated for values of
$\rho_b$ for which the volume phase is still highly diluted. Then the
adsorbed chains form large loops, resulting in an adsorption layer
width which is of the order of the radius of gyration of the polymers
in the bulk.  This effect is well-known for homopolymer
adsorption~\cite{bd,fleer_buch}.  However, increasing $\rho_b$ beyond
the saturation value a different behavior emerges: The chains extend
in the direction perpendicular to the interface, since the loops of
the adsorbed chains begin to stretch in a hairpin-like fashion.  Such
a brush-like regime for multiblock-copolymers at selective interfaces
was recently predicted by Leclerc and Daoud~\cite{ld_mm}. The reason
for this behavior is that the majority blobs (or blocks in
Ref.\cite{ld_mm}) can be squeezed without much loss of interface
energy, since only the alternation of the two blob types across the
interface fixes the chain. In this way more polymer chains can be
adsorbed at the interface.

Our simulation results confirm this picture qualitatively. Distinct
from the behavior of adsorbed homopolymers, we report here for
$\rho_b$ larger than a characteristic value $\rho_b^*$ the stretching
of the adsorbed copolymers in the direction perpendicular to the
interface and their contraction parallel to it.\\

\noindent{\bf II. Simulation algorithm}

Our Monte Carlo simulations for
copolymers were performed using the bond--fluc\-tu\-ation method
(BFM)~\cite{Kremer,Deutsch-Binder}.  The BFM is a lattice algorithm
where each monomer is represented by a lattice cell. Thus on a
three--dimensional simple cubic lattice each monomer occupies eight
neighboring lattice sites. The length of a bond connecting two
neighboring monomers fluctuates between 2 and $\sqrt{10}$ lattice
spacings~\cite{Kremer,Deutsch-Binder}. Self--avoidance (the excluded
volume interaction) is satisfied by not allowing any two monomers to
have a lattice site in common. To avoid bond-crossing the allowed
bonds are restricted to a set of 108
vectors~\cite{Kremer,Deutsch-Binder,HLT_bc}. Here we study the
behavior of $n$ random copolymers, each of length $N$, all placed in a
box of size $L\times L\times H$, with periodic boundary conditions in
the x-- and y--directions and two impenetrable surfaces at $z=0$ and
$z=H$. The copolymers are random; for each of them separately the
$N$-monomer sequence consists of randomly chosen $A$- and
$B$-monomers. For adsorption we assume a symmetrical situation: the
interaction energy of each monomer with its unfavorable solvent is
$\chi k_BT$ and with its favorable solvent is zero. We let the solvent
below the interface ($z \leq H/2$) favor A--type, and the solvent
above the interface ($z \geq H/2+1$) B--type monomers.  Note that the
interface is thus at $z_0=(H+1)/2$. In the Monte Carlo algorithm the
chains move by position changes of their monomers, which attempt
nearest neighbor steps on the underlying cubic lattice. A move is
taken into consideration only if it satisfies the requirements of
self--avoidance and of non--breaking of bonds. Furthermore,
energetically unfavorable moves are statistically permitted according
to the usual prescription involving the Boltzmann factor.
\\

\noindent
{\bf III. Simulation Results}

We study the density dependence of the
adsorption properties by changing $n$, the number of polymers in the
$L \times L \times H$ box. Here we take $L=50$ and $H=100$ and focus
on the results obtained using copolymers of length $N=64$ with a
monomer--solvent interaction parameter of $\chi=3.15$. Results for
other copolymer lengths and for other $\chi$ parameters will be
mentioned when appropriate.  Note that in the single chain case ($n =
1$) the parameters $N=64$ and $\chi=3.15$ let the system be located in
the well adsorbed scaling regime, see Ref.\cite{us_1,us_2}.  

An initial configuration is generated by randomly placing the first
monomer of each polymer in the system and then randomly adding the
subsequent monomers, such that self--avoidance and the restrictions on
the bonds are obeyed. The energetic aspects of the interaction with
the solvents are then taken care of by the usual Boltzmann factor; the
monomer--monomer interaction is only accounted for through the
excluded volume aspect. This means that both solvents are good for
both species.  We established numerically that the relaxation time
(determined using the autocorrelation function of the
radius of gyration $R_g$ and of its z--component $R_{g\perp}$
\cite{Lai}) is around 50,000 Monte Carlo steps (MCS), where a MCS
consists of $nN$ move attempts; we thus view the copolymers as having
reached equilibrium after 200,000 MCS. Averages are then calculated
from the configurations obtained in the subsequent 200,000 MCS. In
order to improve the statistics we average over results from 8
independent runs.

In Figure 1(a) we plot the densities of the A--type monomers
($\rho_A$) and of the B--type monomers ($\rho_B$) as a function of the
height $z$. We pause to make clear how these densities are normalized. 
In the lattice, bond--fluctuation model used here, the maximal number
of monomers which can be accommodated in the given volume is
$\frac{1}{8}L^2H$, since due to the excluded volume restrictions each
monomer blocks 8 lattice sites.  In our system containing $n$ chains
of $N$ monomers each the total number of monomers is $nN$; hence the
average density is $\bar{\rho} = 8Nn/(L^2H)$. Now the densities
$\rho_A$ and $\rho_B$ satisfy the relation
\begin{equation}\label{rho_}
\frac{1}{H} \sum_z (\rho_A(z) + \rho_B(z)) = \bar{\rho}~~.
\end{equation}
The number of polymers in Fig. 1(a) is $n=50$.  Comparing with the
results for single chain, Fig.1 of Ref.\cite{us_1}, we find that now
the monomer densities are still nonzero quite far away from the
interface (i.e. from $z=10$ to $z=30$ and from $z=70$ to $z=90$ in
Fig. 1(a)). Furthermore, at such distances $\rho_A$ and $\rho_B$ are
equal and are independent of $z$. In this range we identify this
constant with the bulk density. When approaching the impenetrable
boundaries at $z=0$ and at $z=100$, $\rho_A$ and $\rho_B$ drop to
zero.  In the following we will not consider the range from $z=0$ to
$z=10$ and from $z=90$ to $z=100$ any further.  Close to the interface
$\rho_A$ and $\rho_B$ peak on their favorable side. The densities
decay smoothly on their favorable side and sharply across the
selective interface, so that their values on the unfavorable side near
the interface lie below the bulk density. This differs from our
findings for single chain, for which we found that the bulk density
is zero and that densities close to the interface on the unfavorable
side display a secondary peak \cite{us_1}. In Fig. 1(b) we show
$\rho_A + \rho_B$ as a function of $z$. The result is a symmetric peak
centered at the interface and superimposed on a bulk density
background.

In Fig. 2(a) we plot the bulk density $\rho_b$ against the average
density $\bar{\rho}$, as given by Eq.(\ref{rho_}). We see that $\rho_b$
becomes extremely low
for large $\chi$ and small $\bar{\rho}$ (but does not disappear completely).
The density $\rho_b$ increases linearly with $\bar{\rho}$ for
larger $\bar{\rho}$. Fig.2(b) shows the density at the interface
$\rho_s=(\rho_A(H/2)+\rho_B(H/2)+\rho_A(H/2+1)+\rho_B(H/2+1))/2$ as a
function of $\bar{\rho}$. For small $\bar{\rho}$, $\rho_s$ increases
rapidly, while for larger $\bar{\rho}$ it increases more slowly, the
cross-over region being around $\bar{\rho} \simeq 0.02$. From the
Figure the almost linear increase of $\rho_s$ in the range of large
$\bar{\rho}$ is also clear.  Fig. 2(c) shows a plot of $\rho_s$ versus
$\rho_b$, where again the two regimes can be seen.  To understand
these regimes we plot in Fig.  2(d) $\rho_s-\rho_b$ against
$\bar{\rho}$.  In terms of Fig.1(b) $\rho_s-\rho_b$ is in fact the
peak's height relative to the background.  We see from Fig. 2(d) that
$\rho_s-\rho_b$ attains its maximum around $\bar{\rho}_c \simeq 0.02$.
For $\bar{\rho} < \bar{\rho}_c$, $\rho_s-\rho_b$ increases with
increasing $\bar{\rho}$ while for $ \bar{\rho} > \bar{\rho}_c$ it
decreases with increasing $\bar{\rho}$. This means that in the low
$\bar{\rho}$ regime ($\bar{\rho} < \bar{\rho}_c$) adding more polymers
to the system leads mainly to an increase in density at the interface,
whereas for higher $\bar{\rho}$ values ($\bar{\rho} > \bar{\rho}_c$)
adding more polymers to the system leads to an overall increase in
density in the bulk. In the second regime both $\rho_s$ and $\rho_b$
increase almost linearly with $\bar{\rho}$, whereas $\rho_s - \rho_b$
decreases roughly linearly with increasing $\bar{\rho}$. Extrapolating
the linear dependence in the second regime of Fig. 2(d) to $\rho_s -
\rho_b = 0$ leads to $\bar{\rho} \simeq 0.95$. We note that a third
regime may exist at still higher $\bar{\rho}$ values, when the
copolymer concentration in the bulk reaches the semi-dilute range, but
investigations in exploring rather dense systems are beyond the scope
of the present paper.

For the record, we like to point out that varying the copolymers'
length, using $N=64$ and $N=128$, we obtained plots almost identical
to those shown in Fig.2. For different $\chi$ parameters ($\chi =
2.20$, $3.15$ and $4.20$), we find that $\rho_b$ is independent of $\chi$
while $\rho_s$ increases with increasing $\chi$.

The difference between $\rho_A$ and $\rho_B$ provides a means to
quantify the interfacial selectivity. Fig. 3(a) displays the data for
$n=50$. We plot $\rho_{A-B} \equiv \rho_A - \rho_B$ for $z \le H/2$
and $\rho_{A-B} \equiv \rho_B - \rho_A$ for $z \ge H/2 +1$; this leads
to a symmetric peak. We find in this density regime ($\bar{\rho}
\simeq 0.1$) that the width of the peak is independent of $N$ but is
controlled by $\chi$. Fig. 3(b) shows how the width of the peak in
Fig. 3(a) varies as a function of $\bar{\rho}$ for different
$\chi$. Here the width is taken as full width at half-height (FWHH),
i.e. the difference $z_2-z_1$, with
$\rho_{A-B}(z_1)=\rho_{A-B}(z_2)=\frac{1}{2} \rho_{A-B}(z_0)$, where
$\rho_{A-B}(z_0)$ is the density at the interface (the peak's height
in Fig. 3(a)). Since our model is based on a lattice, we had to
interpolate between the discrete values, in order to obtain the
(non-integer) FWHH.  We have also evaluated numerically the second
moment of the distribution of Fig. 3(a) and found indications that the
second moment may diverge. In Fig. 3(c) we show the peak's height
$\rho_{A-B}$ as a function of $\bar{\rho}$ and we find that the
relationship is linear above $\bar{\rho}_c$.

Keeping in mind that the interfacial selectivity affects mostly the
chains close to the interface, i.e. the adsorption layer, see
Fig. 3(b), we turn now to a comparative study of the polymers'
behavior around the interface and also away from it. For this we
compute $R_{g\perp}$ and $R_{g\parallel}$, the z-- and the
xy--components of the radius of gyration of each copolymer.  The
center of mass of the $k$th polymer is $\vec{R}_{CM}^{(k)} =
\sum_{i=1}^{N} \vec{R}_{i}^{(k)}/N$, where $\vec{R}_{i}^{(k)} =
(x_i^{(k)}, y_i^{(k)}, z_i^{(k)})$ are the coordinates of the $i$th
monomer within the $k$th chain. For the $k$th chain, $R_{g\perp}^{(k)}
= (\sum_{i=1}^{N} (z_i^{(k)} - z_{CM}^{(k)})^2/N)^{1/2}$, and
$R_{g\parallel}^{(k)} = (\frac{1}{2}\sum_{i=1}^{N} [(x_i^{(k)} -
x_{CM}^{(k)})^2 + (y_i^{(k)} - y_{CM}^{(k)})^2]/N)^{1/2}$. Now we have
to specify which copolymers belong to the space around the interface.
For this we use a $\sigma$-criterion: If the z--component of the
polymer's center of mass is within a distance $\sigma$ from the
interface, i.e. if $|z_{CM}^{(k)} - z_0| \le \sigma$, we view the
polymer as being near the interface, otherwise as being far from it
(here again we disregard the far-off regions close to the fixed
boundaries). Now, evidently $2\sigma$ should be taken larger than the
peak's FWHH. Since the FWHH, Fig. 3(b), depends on $\bar{\rho}$, we
choose $\sigma=10$ in what follows (we checked that the choice of
$\sigma$ does not change the features reported below, provided that
$2\sigma$ is reasonably larger than the FWHH). We now average
$R_{g\perp}^{(k)}$ and $R_{g\parallel}^{(k)}$ for the polymers near
the interface to obtain $R_{g\perp}$ and $R_{g\parallel}$; we do the
same for the polymers far from the interface, which leads to
$\widetilde{R}_{g\perp}$ and $\widetilde{R}_{g\parallel}$. In
Fig. 4(a) and Fig. 4(b) we present plots of $R_{g\perp}$ and
$R_{g\parallel}$ and of $\widetilde{R}_{g\perp}$ and
$\widetilde{R}_{g\parallel}$ as functions of $\bar{\rho}$.

Let us first consider the bulk phase, away from the interface. As can
be readily inferred from Fig. 4(b), $\widetilde{R}_{g\perp}$ is close
to $\widetilde{R}_{g\parallel}$ for all $\bar{\rho}$, which means that
the bulk copolymers' shape is not affected by the interface and that
it is isotropic in space. Note that the large fluctuations of the data
points for small $\bar{\rho}$ values arise from the fact that almost
no chains are located in the bulk until the interface is saturated,
see also Fig. 2(a).

The situation is completely different for polymers near the interface,
Fig. 4(a). For very small values of $\bar{\rho}$, $R_{g\perp}$ is
considerably smaller than $R_{g\parallel}$; hence adsorbed chains tend
to be rather flat. Fig.4(a) shows that $R_{g\parallel}$ decreases
monotonically with increasing $\bar{\rho}$, whereas the opposite is
true for $R_{g\perp}$. This is analogous to the behavior of adsorbed
homopolymers, as discussed by Bouchaud and Daoud~\cite{bd}. Comparing
Fig.4(a) with Fig.2(c) one notes that the chain's extension is, in
contrast to the surface density, a smooth function of $\bar{\rho}$
also at the saturation value $\bar{\rho}_c$.  However, increasing the
density beyond the value $\bar{\rho}^* \simeq 0.05$, where
$R_{g\parallel} = R_{g\perp}$, the extension perpendicular to the
interface gets to be larger than the parallel extension. This is in
accordance with Ref.~\cite{ld_mm}, which predicts (as discussed in the
Introduction) that above a characteristic bulk concentration the loops
of the adsorbed chains will stretch in the direction perpendicular to
the interface, in a hairpin-like fashion. Note that also the chain's
extension parallel to the interface, $R_{g\parallel}$, decreases below
its bulk value $\widetilde{R}_{g\parallel}$ (compare Figs.4(a) and
(b)), i.e. the adsorbed chains begin to get squeezed parallel to the
interface.  Thus varying the bulk concentration only by a small amount
($0< \rho_b < 0.1$) the chain's geometry changes from a flat,
pancake-like shape into a brush-like assembly of stretched loops.
This picture is also supported by the behavior of $\rho_s$ as a
function of $\bar{\rho}$ in the saturated surface regime ($\bar{\rho}
> \bar{\rho}_c$). As can be inferred from Fig.2b, the surface
concentration increases even beyond the saturation threshold
$\bar{\rho}_c$ (which is given in Figs.2 already by the second
numerical point).\\

\noindent
{\bf IV. Conclusions} 

We have investigated the behavior of many
random copolymers in the presence of a selective interface. Previous
studies \cite{Garel,Sommer-Daoud,us_1,us_2} showed that a simple
scaling picture works very well for random copolymer adsorption. In
this paper we have extended our previous work to consider the effects
of sizeable chain concentrations. We found that there are at least two
regimes controlled by the polymer density, as can be inferred from
Figs.2 and Fig.4a.

In the very low density regime (except for the adsorption mechanism)
copolymers at interfaces behave similarly to homopolymers at surfaces. 
Starting from a single chain, which in the adsorbed state is flat, an
increase in the chains' density leads to their crowding at the
interface.  As usual for polymer adsorption, the interface may be
fully covered by chains even when the bulk density is still highly
diluted (compare the surface density $\rho_s$ and the bulk density
$\rho_b$ for the second numerical point in Fig. 2(c)). This can be
easily understood from the fact that the adsorption energy per chain
is in most cases a huge quantity compared to the translational free
energy per chain, i.e. to $kT$.

In the second regime, however, where the interface is already covered,
the only way of adding more chains to it is to squeeze the chains in a
brush-like fashion. As a consequence, $R_{g\perp}$, the radius of
gyration of the adsorbed chains perpendicular to the interface exceeds
the average value $\widetilde{R}_{g\perp}$ in the bulk. A means of
picturing this situation is a brush-like assembly of stretched
loops~\cite{ld_mm}.\\

\noindent
{\bf Acknowledgments}

We are grateful to M. Daoud and T. Ohta for
useful discussions.  This work was supported by the Deutsche
Forschungsgemeinschaft, by the Fonds der Chemischen Industrie, by
PROCOPE administrated by the DAAD, and by the Monbusho Grant--in--Aid,
Japan. G. P. thanks both the Alexander von Humboldt Foundation and the
Japan Society for the Promotion of Science for support during
different stages of the work.

\newpage
\noindent
{\bf Figure Captions}

\noindent
Figure 1(a): Densities of the A--monomers
($\rho_A$) (diamonds) and of the B--monomers ($\rho_B$) (crosses) at
the height $z$. Here the number of polymers is $n=50$, the length of
all polymers is $N=64$, and the monomer--solvent interaction parameter
is $\chi=3.15$. \\

\noindent
Figure 1(b): Display of $|\rho_A+\rho_B|$ vs. $z$, see Fig.1(a) for
definitions.\\

\noindent
Figure 2(a): The bulk density ($\rho_b$) vs. the average density
($\bar{\rho}$), see text for details. \\

\noindent
Figure 2(b): The density at the interface ($\rho_s$)
vs. $\bar{\rho}$.\\
 
\noindent
Figure 2(c): $\rho_s$ vs. $\rho_b$. \\

\noindent
Figure 2(d): $\rho_s - \rho_b$ vs. $\bar{\rho}$.\\

\noindent
Figure 3(a): $\rho_{A-B}$ (the absolute value of $\rho_A - \rho_B$)
vs. the height $z$. Here $n=50$. \\

\noindent
Figure 3(b): The FWHH of Fig.3(a) vs. $\bar{\rho}$ for three different
$\chi$ parameters, $\chi=2.20$ (diamonds), $\chi=3.15$ (crosses),
$\chi=4.20$ (squares).\\

\noindent
Figure 3(c): $\rho_{A-B}(z_0)$ vs. $\bar{\rho}$, with $z_0$ being the
interface's location.\\

\noindent
Figure 4(a): Radius of gyration for polymers near the interface:
$R_{g\perp}$ (diamonds) and $R_{g\parallel}$ (crosses)
vs. $\bar{\rho}$.\\

\noindent
Figure 4(b): Radius of gyration for polymers away from the interface:
$\widetilde{R}_{g\perp}$ (diamonds) and $\widetilde{R}_{g\parallel}$
(crosses) vs. $\bar{\rho}$.\\

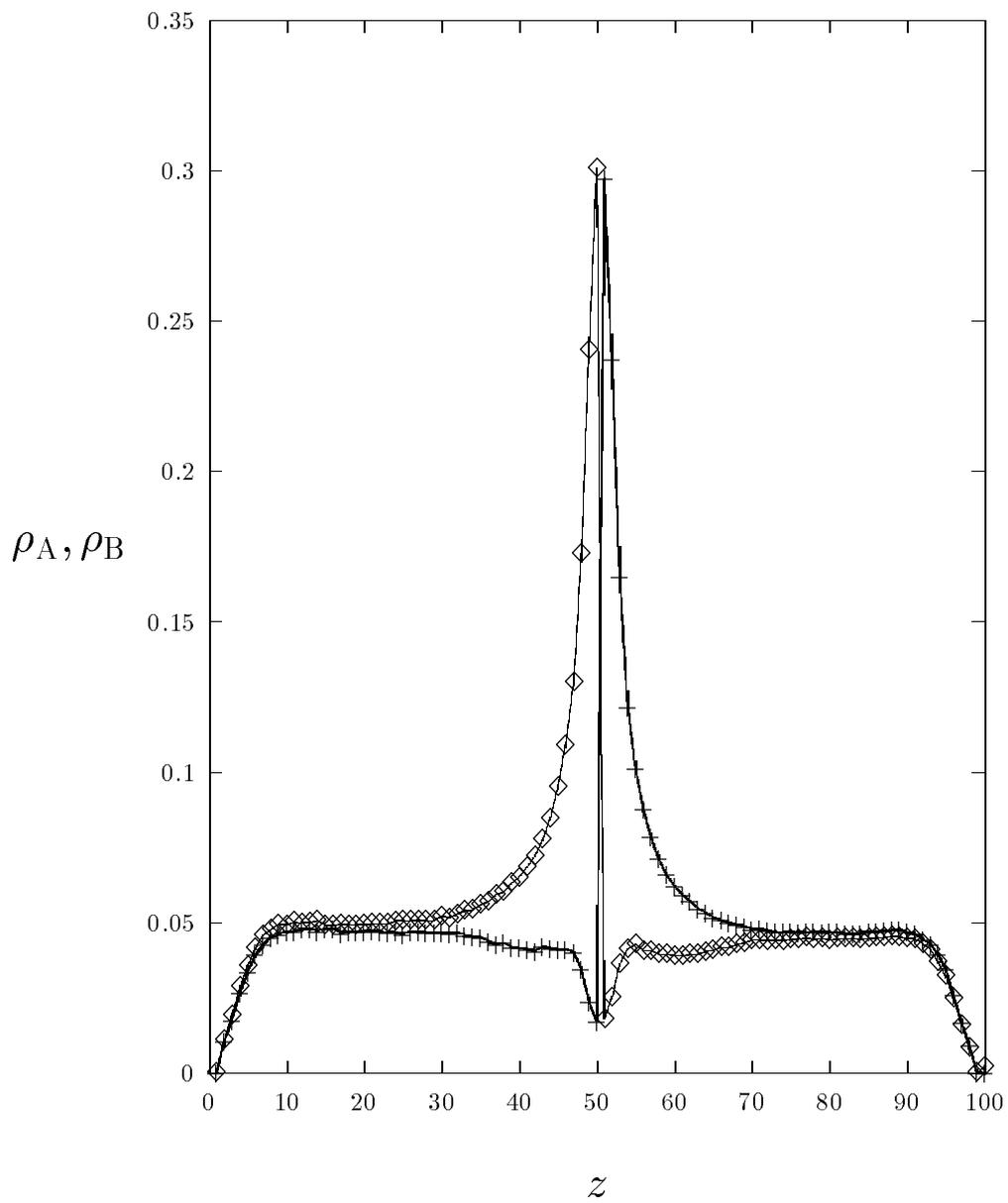
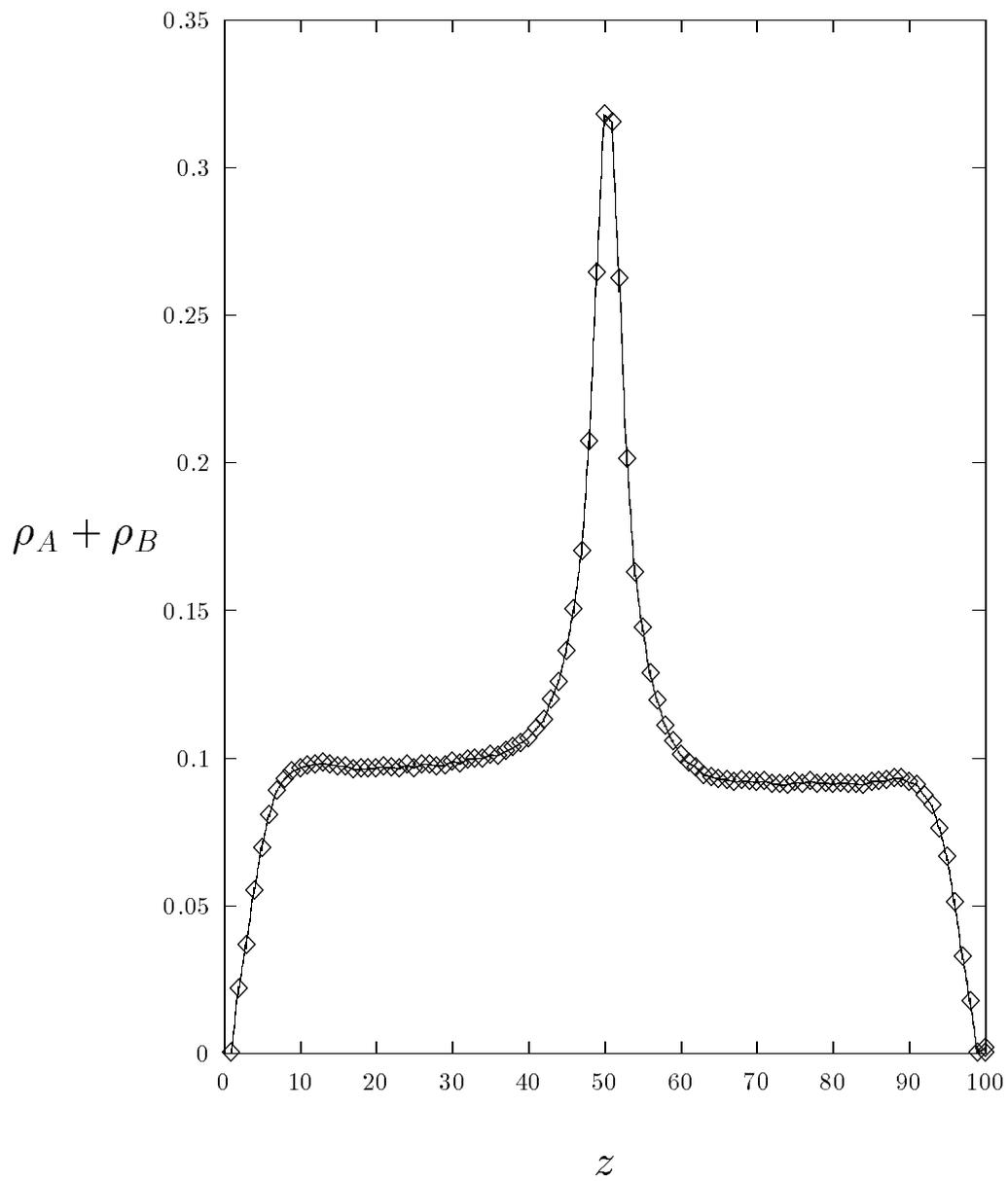
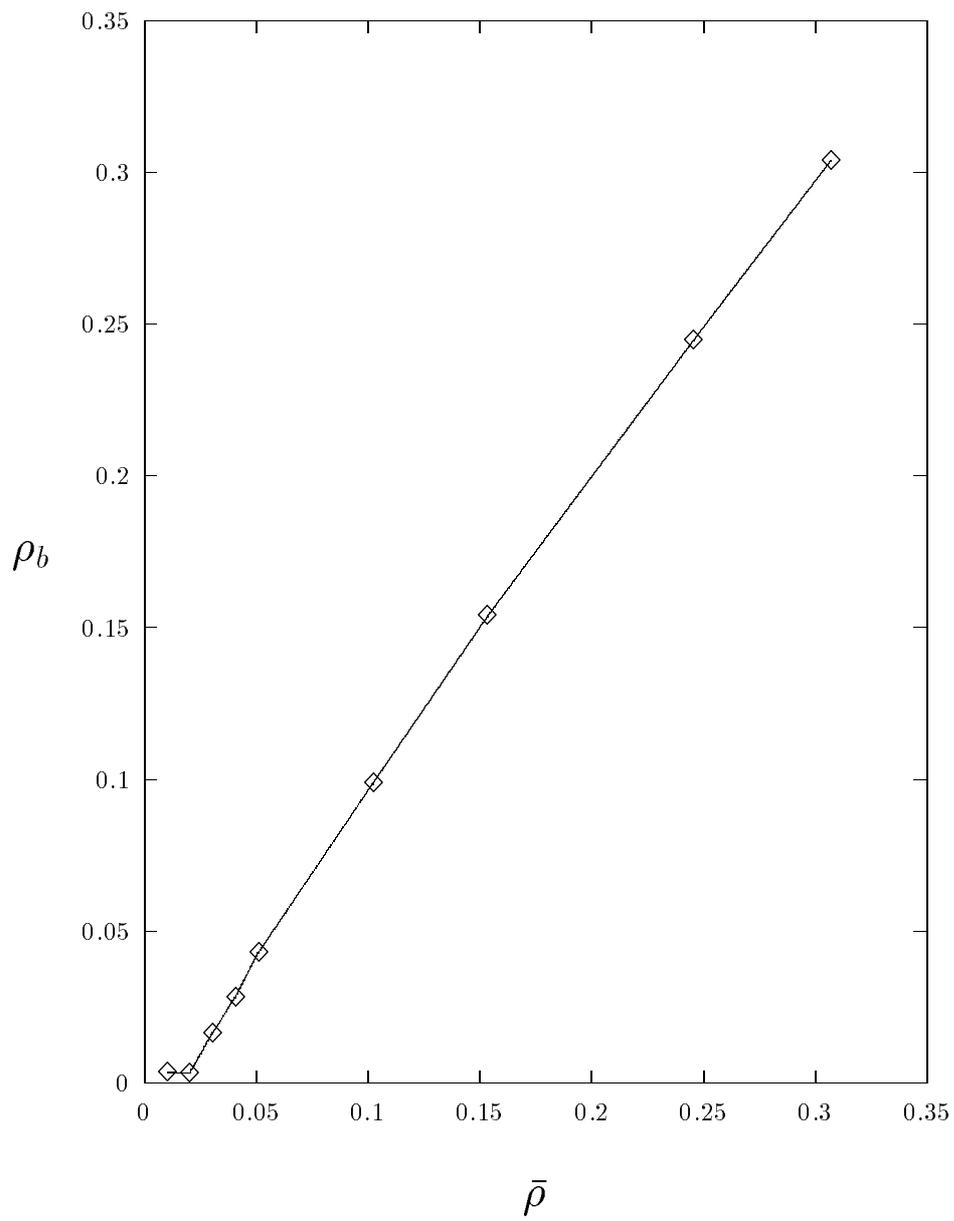
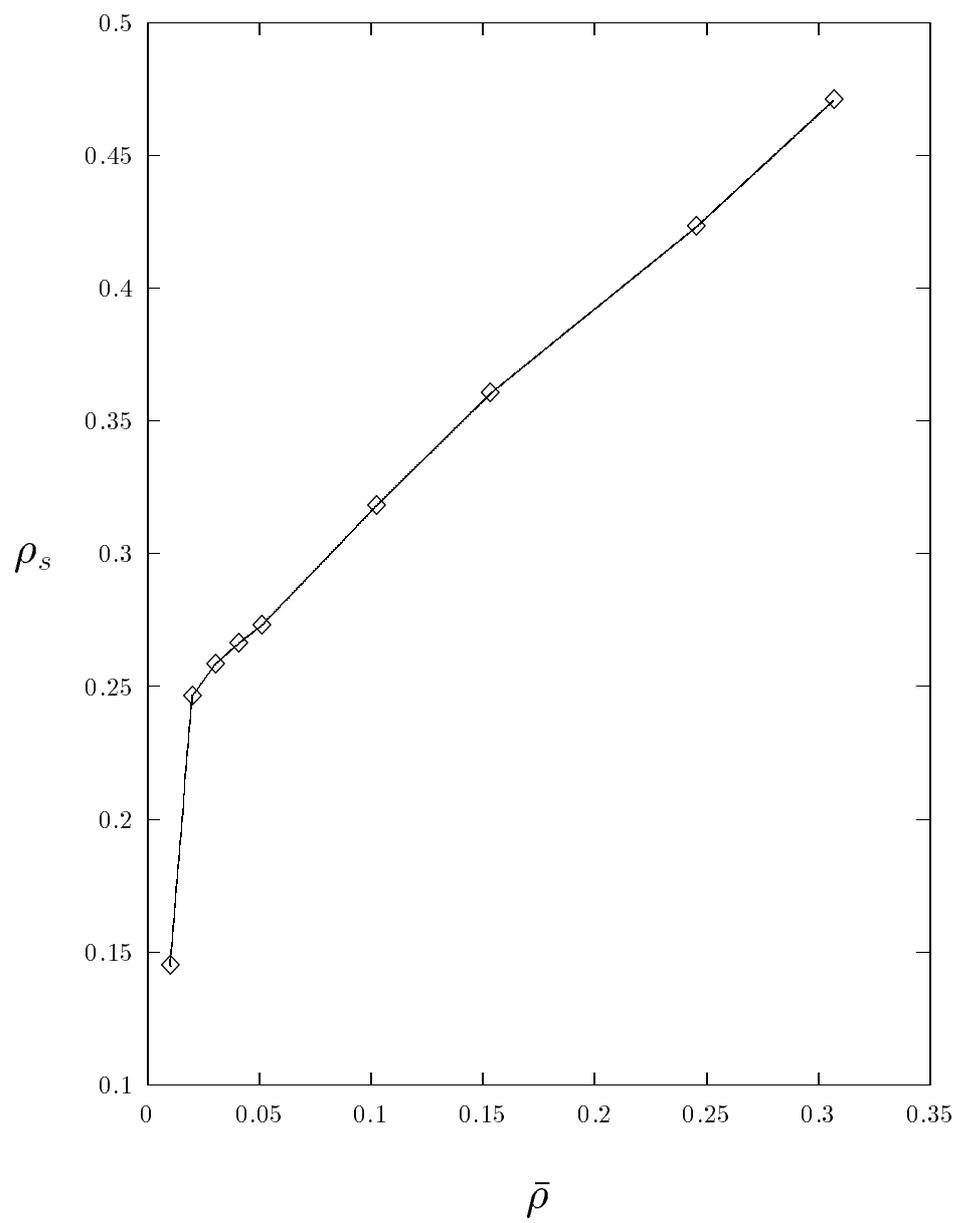
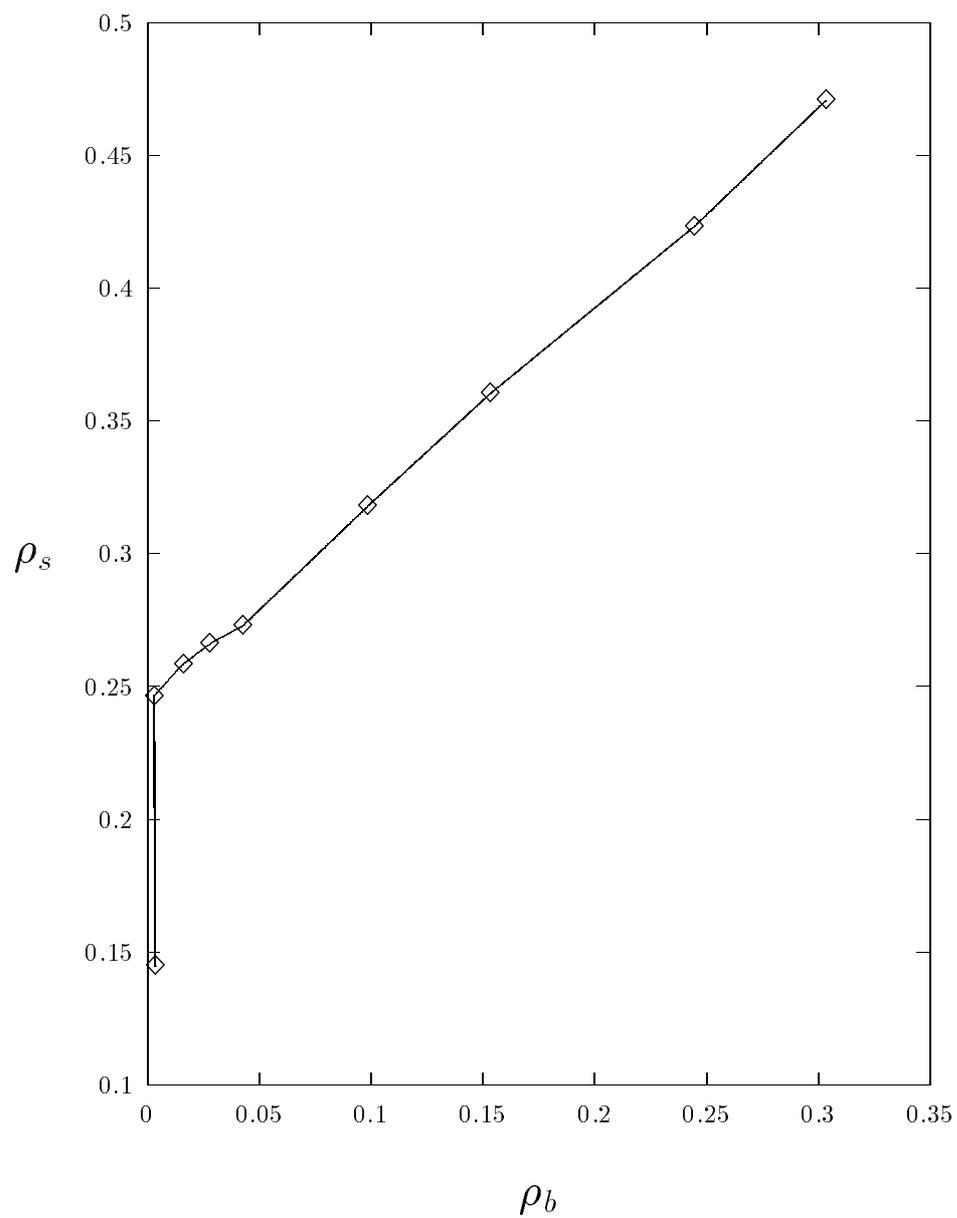
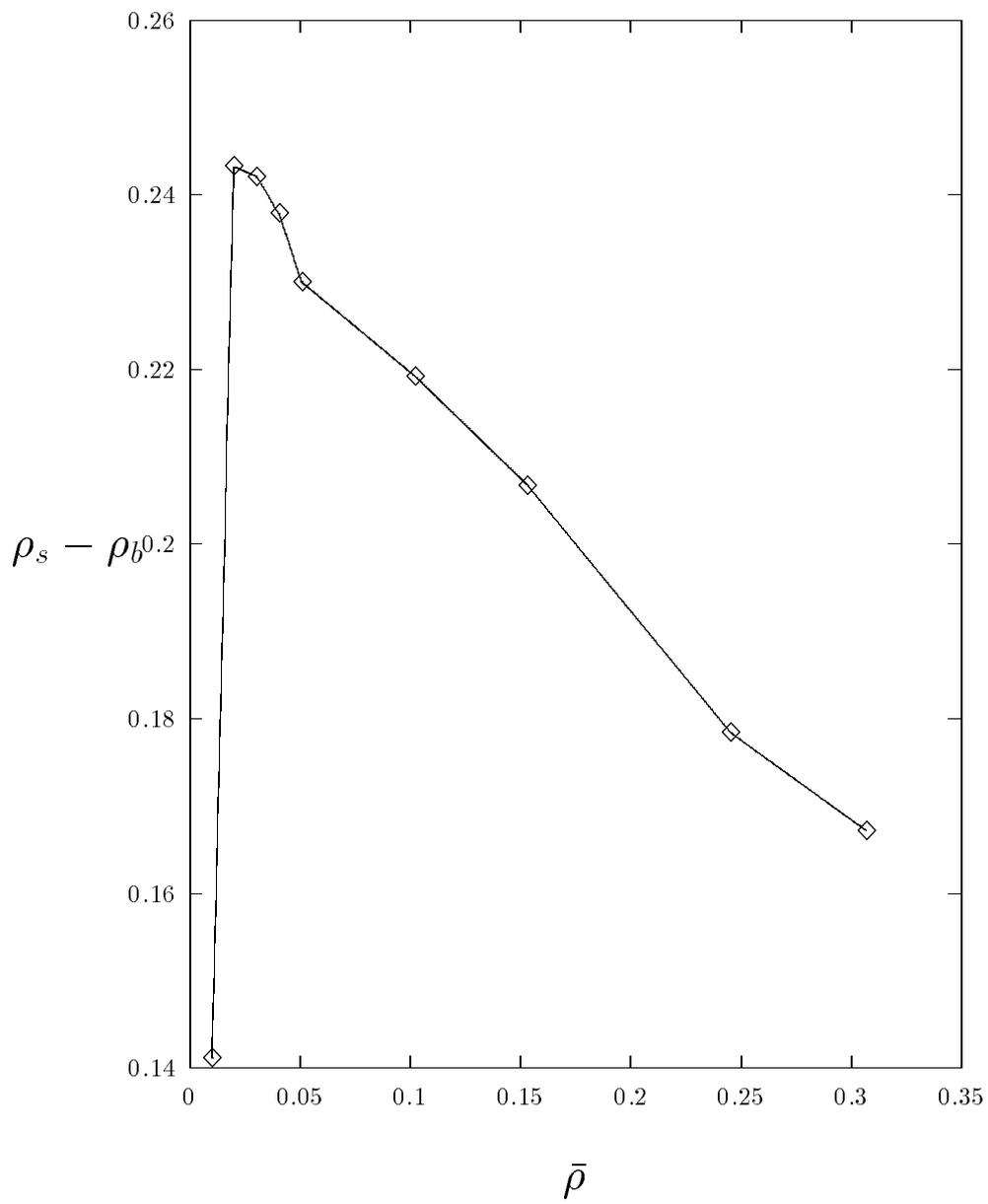
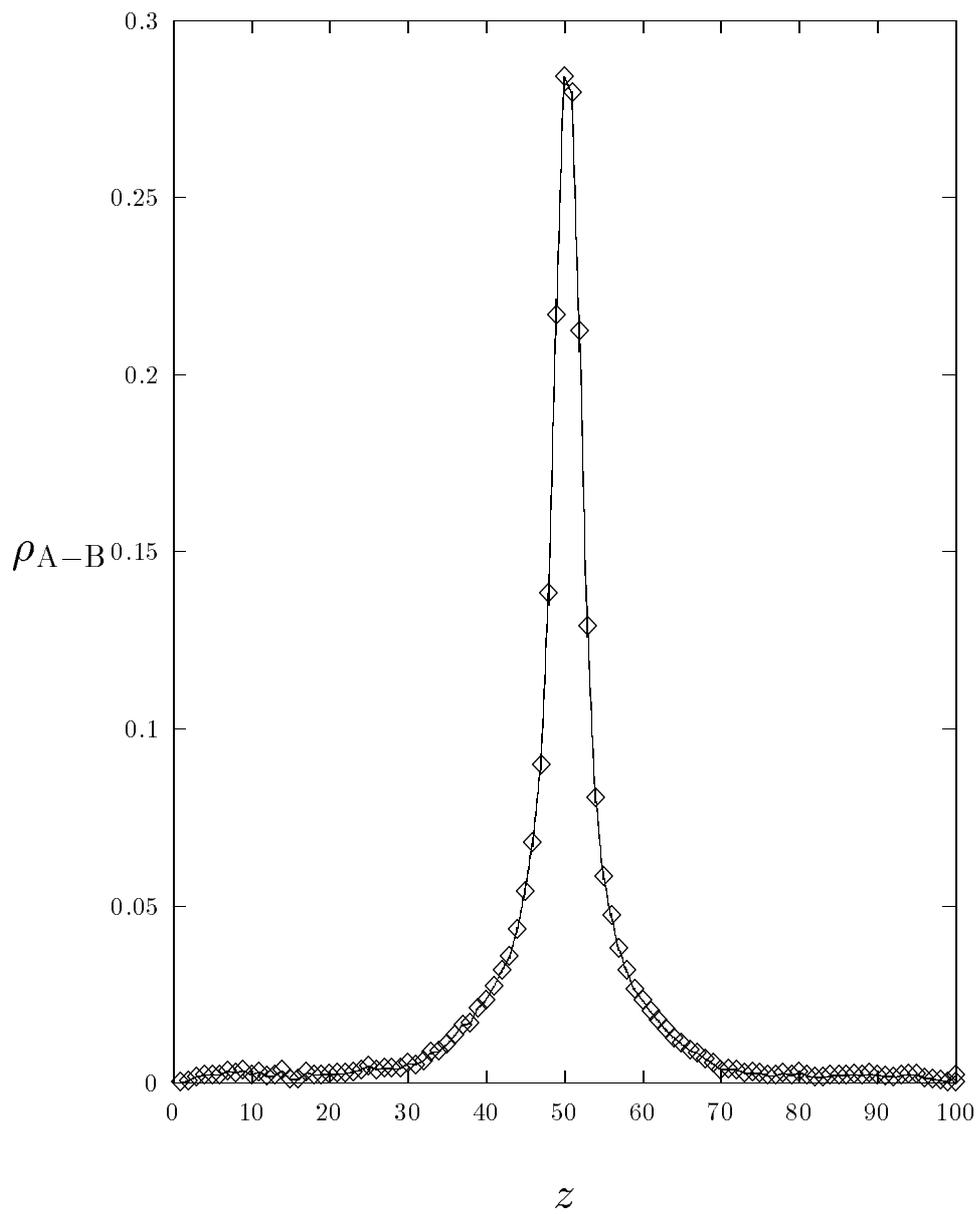
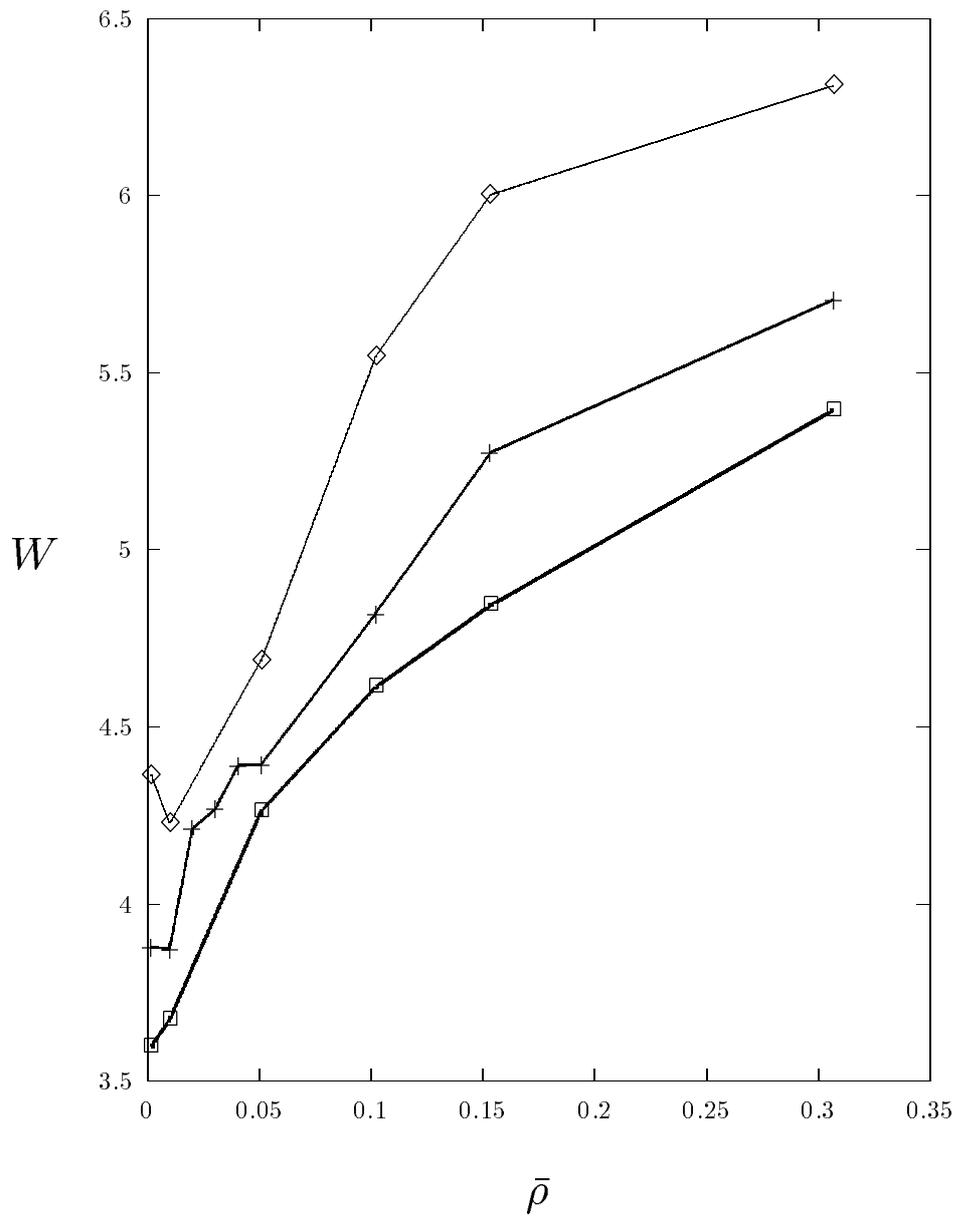
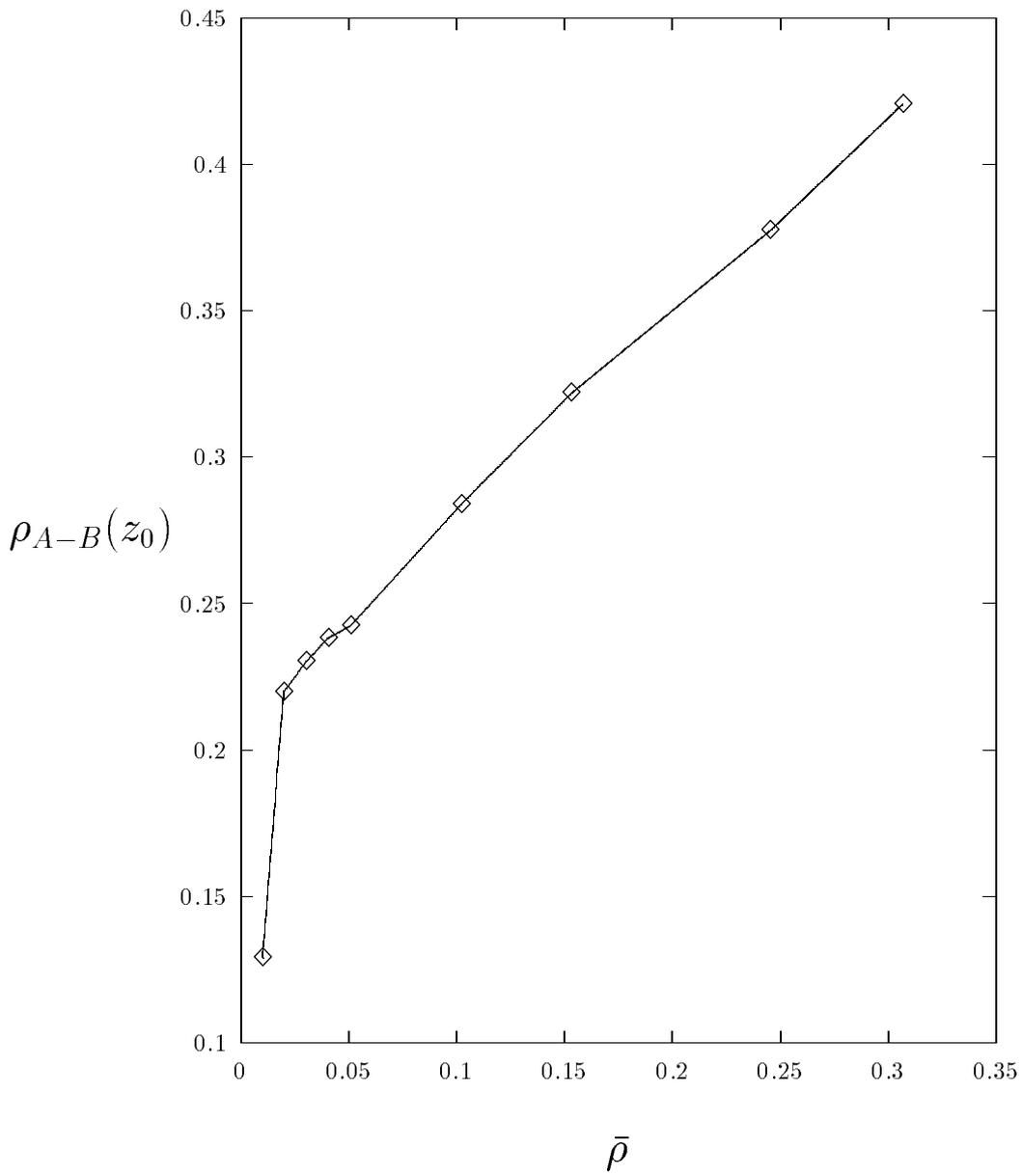
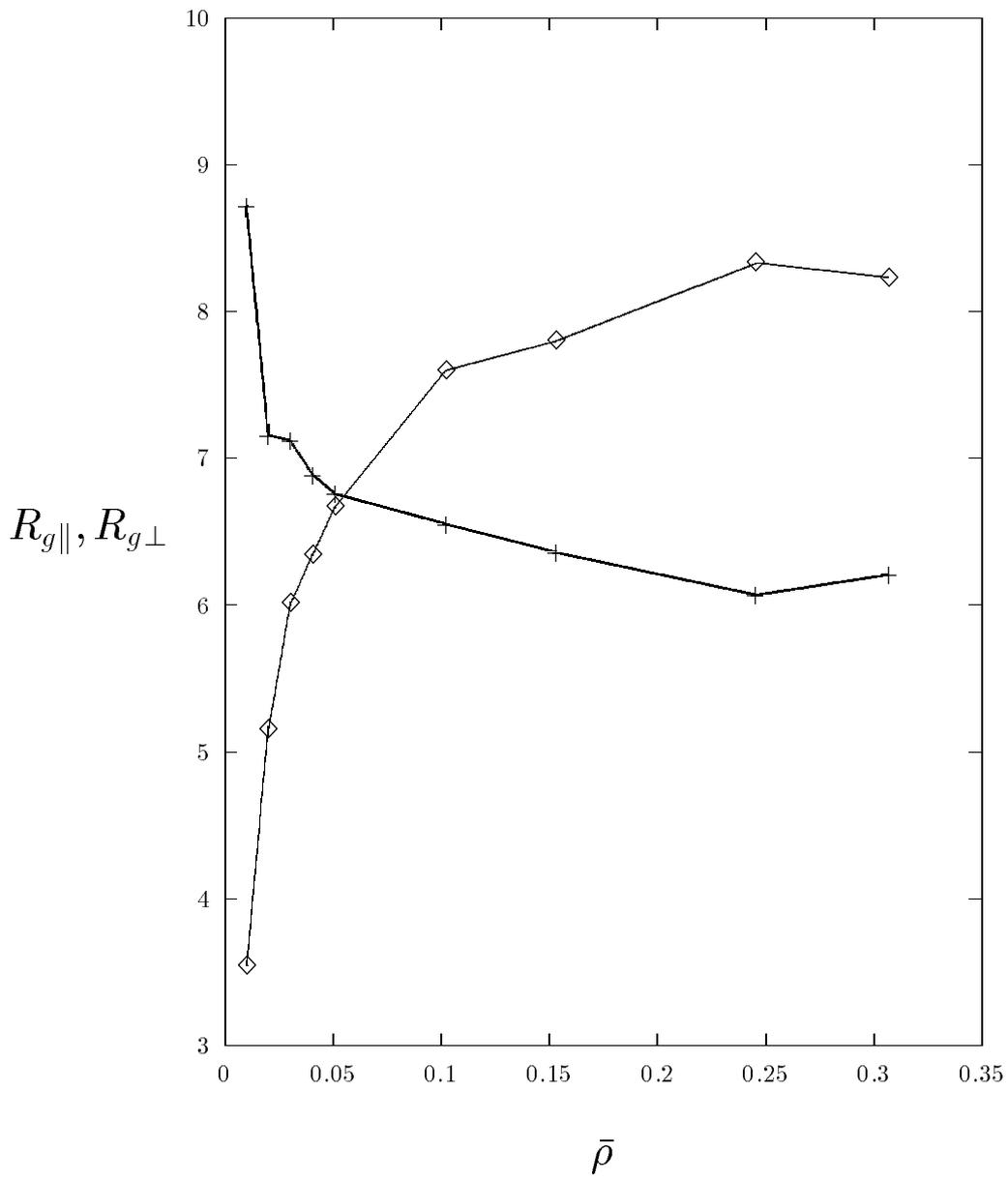
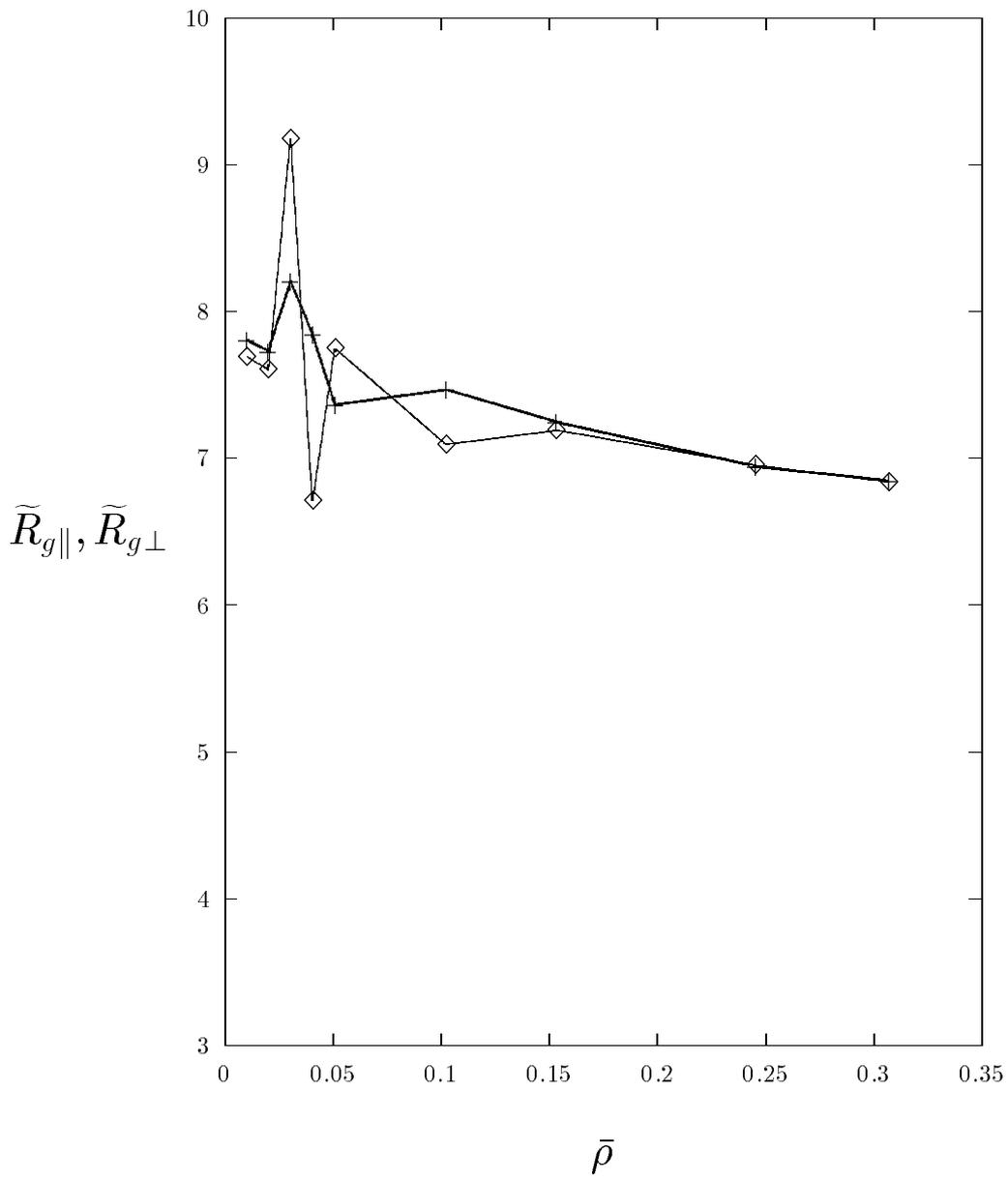
\end{document}